# Quantum Nature of Plasmon-Enhanced Raman Scattering

**Authors:** Patryk Kusch[1], Sebastian Heeg[1,2], Christian Lehmann[1], Niclas S. Müller[1], Sören Wasserroth[1], Antonios Oikonomou[3], Nicholas Clark[2], Aravind Vijayaraghavan[2], Stephanie Reich[1,*]

[1] Department of Physics, Freie Universität Berlin, 14195 Berlin, Germany.

[2] School of Materials, University of Manchester, Manchester M13 9PL, United Kingdom.

[3] National Graphene Institute, University of Manchester, Manchester M13 9PL, United Kingdom.

*Correspondence to: stephanie.reich@fu-berlin.de

**TOC graphic**

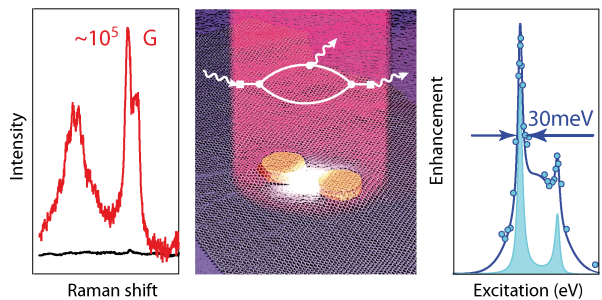

## Abstract

We report plasmon-enhanced Raman scattering in graphene coupled to a single plasmonic hotspot measured as a function of laser energy. The enhancement profiles of the G peak show strong enhancement (up to $10^5$) and narrow resonances (30 meV) that are induced by the localized surface plasmon of a gold nanodimer. We observe the evolution of defect-mode scattering in a defect-free graphene lattice in resonance with the plasmon. We propose a quantum theory of plasmon-enhanced Raman scattering, where the plasmon forms an integral part of the excitation process. Quantum interferences between scattering channels explain the experimentally observed resonance profiles, in particular, the marked difference in enhancement factors for incoming and outgoing resonance and the appearance of the defect-type modes.

Plasmons are the collective excitations of free electrons in metals. The excitation of plasmons in metallic nanostructures produces intense and strongly localized near fields that enhance light-matter interaction.[1,2] Particularly striking is the plasmonic enhancement in light scattering.[3,4,5,6] Surface-enhanced Raman scattering (SERS) enables single-molecule detection and tip-enhanced scattering (TERS) allows Raman imaging with a resolution below the diffraction limit.[7,8,9,10,11] The development of advanced nanoscale fabrication techniques enabled the tailoring of plasmonic near fields for a desired enhancement.[12,13] Such rationally designed hotspots are prime systems for enhancing absorption, luminescence and photoconductivity for light harvesting, optical communication, and ultrasensitive light detectors.[2,14,15]

Well-defined plasmonic hotspots can also be used for fundamental tests of enhanced optical processes in plasmon-probe systems. A full understanding of the plasmonic enhancement in inelastic light scattering requires wavelength-resolved experiments on well characterized hotspots and probe systems for the Raman effect. Graphene is an excellent probe for plasmon-enhanced Raman scattering (PERS).[16,17,18,19] It is a two-dimensional material that interacts with a large fraction of the plasmonic near field. The Raman spectrum of graphene is well established experimentally and understood theoretically.[20,21,22] Most importantly, the Raman intensity of the *G* peak in graphene is independent of laser energy and polarization. Changes in intensity close to plasmonic nanostructures are induced by the near field as demonstrated by us for graphene[17,19] and carbon nanotubes.[23,24]

In this letter we measure and analyze strong resonances in the plasmon-enhanced Raman spectrum of graphene that are induced by exciting the localized surface plasmon of a gold nanodimer. The plasmon-induced resonances are narrow in energy (30-50 meV full width at half maximum) and provide up to $10^5$ enhancement. The incoming and outgoing resonances differ strongly in intensity, which contradicts the conventional electromagnetic enhancement model.[25,26] We propose a quantum mechanical description of plasmon-enhanced Raman scattering, which excellently describes the experimental observations. Our theory predicts defect-type Raman scattering in perfect graphene when exciting at the localized surface plasmon resonance (LSPR). We experimentally verify this striking consequence of the quantum nature of PERS.

The individual plasmonic hotspot used in this study is shown schematically in Fig. 1a. It consists of a lithographically fabricated gold dimer constructed by two nanodiscs that are separated by a small gap. One of the dimer plasmon modes will produce a very intense near field that is confined to the dimer cavity (glow in Fig. 1a). We showed that such dimers enhance Raman scattering in graphene and carbon nanotubes.[17,19,23,24] Gold dimers with 110nm disc diameters and 20nm gaps were prepared by electron-beam lithography on a Si/SiO$_2$ substrate (300nm layer of thermally grown SiO$_2$). The dimers were exposed by an electron-beam in a LEO 1530 Gemini FEG SEM and their Cr/Au (5/40 nm) metallization was performed by an electron-beam evaporation system. A micromechanically exfoliated graphene flake was transferred onto the dimer structures (Fig. 1a) using a polymer support layer that was subsequently dissolved.[17] Accurate placement of the graphene membrane on the dimers was achieved using a transfer system manufactured by Graphene Industries. Characterization of the dimers by scanning electron microscopy (SEM), atomic-force microscopy (AFM), dark-field and Raman spectroscopy are given in the Supplementary Information; further details can be found in Ref. 17.

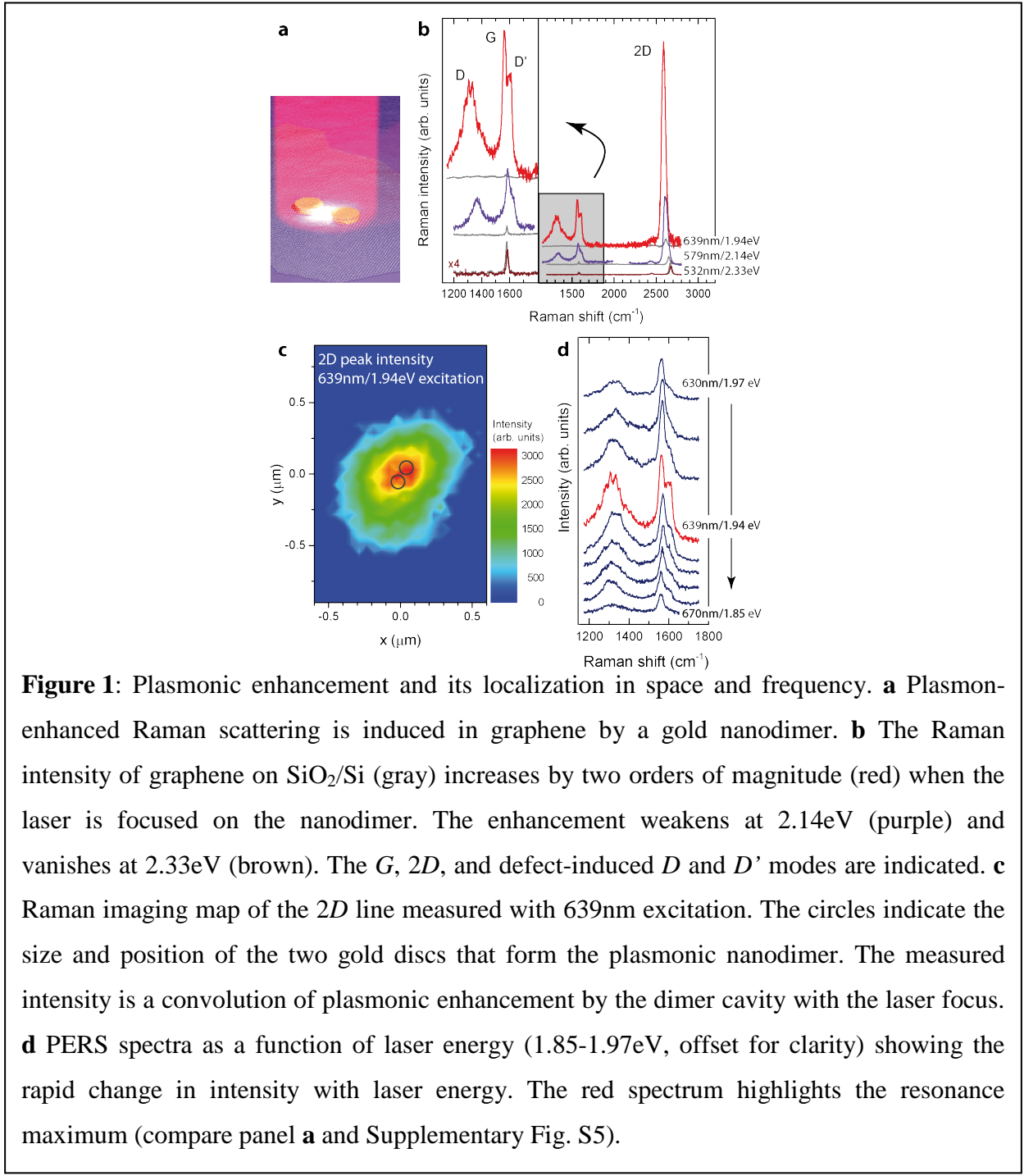

**Figure 1**: Plasmonic enhancement and its localization in space and frequency. **a** Plasmon-enhanced Raman scattering is induced in graphene by a gold nanodimer. **b** The Raman intensity of graphene on $SiO_2$/Si (gray) increases by two orders of magnitude (red) when the laser is focused on the nanodimer. The enhancement weakens at 2.14eV (purple) and vanishes at 2.33eV (brown). The *G*, *2D*, and defect-induced *D* and *D'* modes are indicated. **c** Raman imaging map of the *2D* line measured with 639nm excitation. The circles indicate the size and position of the two gold discs that form the plasmonic nanodimer. The measured intensity is a convolution of plasmonic enhancement by the dimer cavity with the laser focus. **d** PERS spectra as a function of laser energy (1.85-1.97eV, offset for clarity) showing the rapid change in intensity with laser energy. The red spectrum highlights the resonance maximum (compare panel **a** and Supplementary Fig. S5).

In this work we focus on the dependence of plasmonic enhancement in Raman scattering on excitation energy, which required fully tunable laser and detection systems. Raman scattering was excited with a dye laser for excitation wavelengths 565-600nm and 620-680nm; for longer wavelengths a Ti:Sa laser was used (675nm and higher). Spectra were recorded every 2 nm close to resonances and 5 nm further away from the resonances. The power of the incoming laser was ≈200 µW. Elastically scattered light was suppressed by edge filters; the inelastically scattered light was dispersed by a single-grating Horiba T64000 spectrometer and detected by a CCD. The laser was focused onto the nanodimer-graphene sample with a 100x objective; the focal diameter was 700 nm as obtained by measuring the *G* peak intensity over the edge of freely-suspended graphene. The laser focus was carefully centered on the nanodimer using an XYZ nanopositioning piezo stage. Raman imaging maps were obtained by varying the position of the piezo stage in steps of 50nm and recording Raman spectra at each spot. Some Raman imaging maps were also obtained on an Horiba XploRa Raman spectrometer because of its smaller spot size (570 nm).[17]

The plasmon-enhanced Raman spectra were normalized by the *G* and 2*D* peak intensity measured on freely-suspended graphene. We used freely-suspended graphene as a reference, because the Raman intensity of graphene on Si/SiO$_2$ changes with excitation energy due to Fabry-Perot-type interferences on the graphene-SiO$_2$-Si structure.[27] The Raman spectra were fit by Lorentzian line shapes; exemplary fits are shown in Supplementary Fig. S5. Enhancement profiles were obtained by plotting the Lorentzian areas as a function of excitation energy.

Figure 1b compares the Raman spectra when the laser focus was on the dimer (red, purple, and brown) and away from the dimer (gray). For excitations at 1.94 and 2.14 eV we observe strong plasmonic enhancement, i.e., the red and purple spectra are much more intense than the gray reference. The enhancement vanished at 2.33 eV excitation; the brown and gray curve are identical.[17] The enhanced spectra differ in three characteristic features from the reference signal: First, the overall intensity of the Raman peaks increases dramatically. The *G* peak is enhanced up to a factor of 100 when the laser focus is moved on the nanodimer (spectrum at 1.94 eV in Fig. 1b). In Fig. 1c we present a Raman imaging map of the integrated 2*D* line intensity at this energy; the map shows a single hotspot in the nanocavity that is folded with the focus of the laser (diameter ~700nm). Already a shift of the focus by 50 nm led to a 10% decrease in scattering intensity, which is a remarkable sensitivity to position. Converting the measured increase in total intensity into the enhancement of the Raman cross section is challenging for a two-dimensional Raman probe, because the area and intensity distribution of the plasmonic near field are not known.[17] Single molecule probes as used in wavelength-dependent SERS experiments on individual hotspots are advantageous in this respect.[28] An order-of-magnitude estimate for the enhancement in the cross section is obtained by considering that only a fraction of $10^{-3}$-$10^{-4}$ of the laser focus overlaps with the dimer nanocavity and contributes to the plasmon-enhanced Raman spectrum. The increase in total intensity by $10^2$ (Fig. 1b), therefore, corresponds to an enhancement of $10^5$-$10^6$ in the Raman cross section, see Ref. 17 and the Supplementary Information for further discussion. The plasmonic enhancement by our rationally designed lithographic hotspot is approaching the maximum $10^7$-$10^9$ fold enhancement of single-molecule SERS.[29] It is higher than the $10^2$-$10^4$ fold enhancement typical of TERS.[10]

The second difference between the spectra with and without plasmonic enhancement is that the *G* and 2*D* peaks in the enhanced spectra are lower in frequency than the reference peaks taken away from the nanodimer (Fig. 1b). This comes from a tensile strain in the transferred graphene flakes on top of the gold dimers. The strain shifts the phonon frequencies and allows to discriminate between enhanced spectra and the background scattering by graphene based on the phonon frequency.[17,19] This is a characteristic feature of our graphene/nanodimer system. When exciting with a laser energy outside the plasmonic resonance the intensity of the phonons from strained graphene is zero, Fig.1b and Supplementary Fig. S5, because the strain is localized in a small area around the dimer.[17,19]

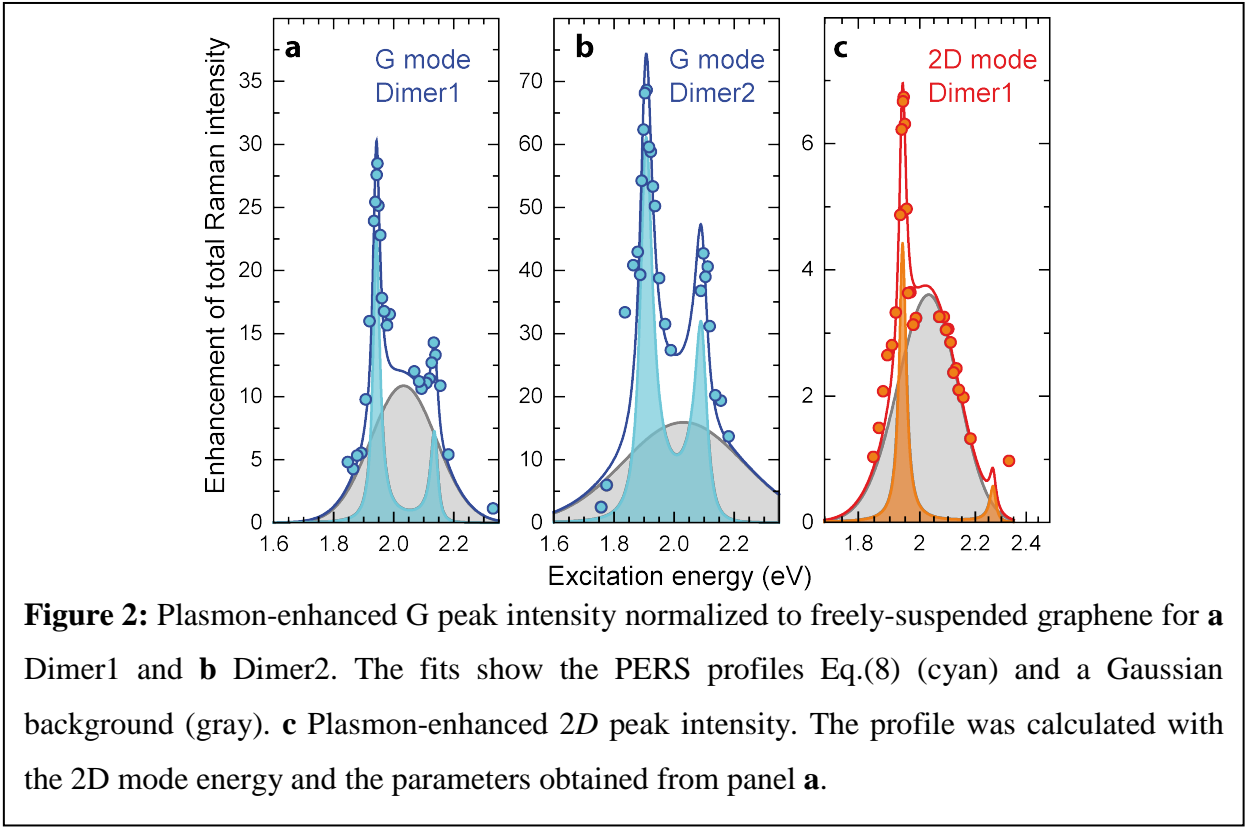

**Figure 2:** Plasmon-enhanced G peak intensity normalized to freely-suspended graphene for **a** Dimer1 and **b** Dimer2. The fits show the PERS profiles Eq.(8) (cyan) and a Gaussian background (gray). **c** Plasmon-enhanced 2D peak intensity. The profile was calculated with the 2D mode energy and the parameters obtained from panel **a**.

The third difference between the spectra in spatial and energetic proximity of the plasmon and away from it is the appearance of the defect-induced *D* and *D'* peaks by plasmonic enhancement. The PERS spectra in Fig. 1b and d show strong *D* and *D'* modes with intensity ratio $I_D/I_G=3$ and $I_{D'}/I_G=1.2$ at 1.94eV. Usually, scattering by these modes is a sign of defects in the graphene lattice.[20,21,30] Defects, however, are not the source of the *D* and *D'* peaks in our experiment: $I_D/I_G$ resonates strongly with the LSPR and vanishes for off-resonant excitation, Fig. 1b. The *D* peak also disappears if the incoming and scattered light are polarized perpendicular to the dimer axis [(⊥,⊥) scattering configuration, Supplementary Fig. S8], although the *G* peak enhancement only decreased by a factor of five. Also, mechanically exfoliated graphene typically has a high crystal quality with no detectable *D* mode signal.[21] We will show below that defect-mode scattering is explained by the quantum theory of PERS proposed in this paper.

The plasmonic enhancement induced by the nanodimer depends strongly on excitation energy. Starting at 1.85eV excitation energy (Fig. 1d) the enhancement reaches a first maximum at 1.94eV excitation. It then drops dramatically for further increasing laser energy, but reaches a second maximum at 2.14eV (Fig1b). Finally, for green excitation there is no detectable spectrum from strained graphene, see 2.33eV spectrum in Fig. 1b. When plotting the Raman intensity of the *G* peak as a function of laser energy (enhancement profile, Dimer1) in Fig. 2a we find a pair of narrow resonances (1.94 and 2.14eV) superimposed on a broad resonant background (2.03eV). Figure 2b shows the enhancement profile of a second dimer Dimer2. Again we find two well-defined resonances on a broad background enhancement.

The two resonance maxima in Fig. 2a and b have an energetic separation that matches the energy of the *G* phonon (200meV). They correspond to an incoming resonance with the LSPR, i.e., the laser matches the LSPR, and an outgoing resonance, i.e., the scattered (outgoing) photon matches the LSPR. This interpretation is confirmed by the enhancement profile of the *2D* peak (phonon energy 320meV) induced by Dimer1 (Fig. 2c). Its incoming resonance matches the 1.94 eV resonance of the *G* peak. The outgoing resonance is moved to higher excitation energy (predicted for 2.26 eV); unfortunately, we lack tunable lasers in this energy range to observe the outgoing *2D* resonance. Clearly resolved pairs of incoming and outgoing resonances have never been reported in plasmon- and surface-enhanced Raman spectroscopy. They were not resolved in experiments that average over multiple hotspots, because every hotspot provided a different resonance energy yielding an inhomogeneously broadened profile.[31,32] In a single-molecule SERS experiment on an individual hotspot no outgoing resonance was detected due to the lack of fully tunable excitation.[28]

The narrow resonance peaks in the PERS enhancement profiles, cyan and orange traces in Fig.2, originate from exciting a plasmon that produces a highly localized near field in the nanocavity, Fig.1a. A line scan across the dimer recorded at the maximum of the incoming resonance, Supplementary Fig. S6, has a width (570±5)nm that is identical to the focal diameter of the laser (570±7)nm meaning that the source of the scattering is much smaller than the laser spot. The resonant background shown in gray in Fig. 2 is also due to plasmonic enhancement. It originates from plasmon-induced near fields that occur over the entire surface of the discs forming the dimer. To verify this interpretation we performed a line scan at the maximum of the gray profile, Supplementary Fig. S6. The width of the profile increased to (740±20)nm, which is the width expected for scattering by the two independent gold discs.[17] At first sight the ratio between the maximum nanocavity enhancement at the incoming resonance $I_{in}$ and the background dimer enhancement $I_{bg}$ appears to be small, $I_{in}/I_{bg}=3$ for Dimer1 and 6 for Dimer2. However, $I_{in}$ originates from an area on the order of 20x20nm$^2$ (characteristic size of the cavity), which is much smaller than the entire dimer with an area 200x100nm$^2$ giving rise to $I_{bg}$. The difference between the enhancement in the Raman cross section of the resonance peaks and the background is thus at least two orders of magnitude, which implies that the background has an enhancement ~10$^3$. We note that the simultaneous observation of the nanocavity resonance and the delocalized dimer resonance in a single enhancement profile, Fig. 2, is characteristic of two-dimensional probes like graphene. Here the entire area of the near field contributes to the total intensity, which we call a hotspot-dominated system. In single-molecule SERS experiments the size of the probe (molecule) determines the overall intensity for a given enhancement per unit area (probe-dominated system). Two-dimensional and molecular probes complement each other in studying the fundamentals of plasmonic enhancement in light scattering.

We now come back to the incoming and outgoing resonances with nanocavity plasmons, which will be the focus of the remaining part of the paper. We will refer to these resonances as a PERS profile for simplicity (cyan and orange traces in Fig.2, compare also the background-substracted profiles in Fig.4). The narrow resonances have a full width at half maximum (FWHM) of 28 meV in Fig. 2a (corresponding to 9nm at the resonance wavelength) and 50 meV in Fig. 2b. The resonance width is an order of magnitude smaller than the peak in elastic scattering as measured with dark-field spectroscopy (Supplementary Fig. S3). It is a surprisingly large ratio, although some difference between the cross section for inelastic (Raman) and elastic (Rayleigh or dark field) scattering is expected.[25,28,33] The reason is that the elastic cross section is determined by the entire induced dipole of the nanodimer regardless of the near-field distribution, whereas plasmonic enhancement in Raman scattering scales with the localization of the near field.[33] To best of our knowledge there are only two previous studies reporting resonance profiles of SERS enhancement from a single hotspot.[28,34] Both observed resonances with similar widths as found by us (FWHM 50 meV in Ref. 28 and 50 and 80 meV in Ref. 34). No pairs of incoming and outgoing resonance were detected in these studies due to limitations in tunable excitation and detection (Supplementary Information). In view of these collected data narrow resonances appear to be characteristic for strong plasmonic enhancement in inelastic light scattering, although this has not been realized before. An interesting aspect is that narrow resonances are easily missed in Raman experiments with a single or a few laser lines.[7,31,35] It explains why rationally designed nanostructures were mistakenly argued to be poor systems for plasmonic enhancement compared to rough metal surfaces. This is clearly not the case given the strong incoming resonance induced by the nanodimer. Rough surfaces harbor a large number of hotspots with varying LSPR. At fixed laser energies some of the hotspots will match the resonance conditions. Controlled nanofabrication, in contrast, will lead to uniform hotspots with tailored characteristics in enhancement, energy, width, and polarization.

We now analyze the characteristics of the PERS profiles in Fig.2. The outgoing resonances in Fig. 2a and b are weaker than the incoming resonances. The difference in maximum intensity of the two resonances is an uncommon feature in resonant Raman scattering;[36] it contradicts the electromagnetic enhancement (EM) model that is commonly used to explain PERS.[7,13,16,25] The EM model describes the plasmon as an external antenna that increases light absorption and emission during Raman scattering. The presence of the LSPR leads to an increase in the electromagnetic field by an enhancement factor $g(\hbar\omega)$, where $\hbar\omega$ is the photon energy. The Raman intensity of a phonon *ph* within the EM model is given by [1,25,37]

$$I_{ph}^{EM} \propto \left|K_{ph}^{EM}\right|^2 = g(\hbar\omega)^2 g(\hbar\omega - \hbar\omega_{ph})^2 \left|K_{ph}^{Raman}\right|^2, \tag{1}$$

where $K_{ph}^{Raman}$ is the Raman matrix element and $K_{ph}^{EM}$ the matrix element for plasmon-enhanced Raman scattering within the EM approximation. Often Eq.(1) is further simplified by neglecting the phonon compared to the photon energy. It leads to a $g(\hbar\omega)^4$ expression of the plasmonic enhancement (commonly referred to as the $E^4$ scaling of SERS).[25,33] We stress that this approximation is not valid in our system, because the resonance width (50 meV) is small compared to the phonon energy (200 meV *G* and 320 meV *2D* line). Equation (1) is symmetric in the incoming $\hbar\omega_1$ and outgoing channel $\hbar\omega_2$, i.e., the two resonances must be of equal intensity in contrast to the experimental findings.

Inelastic light scattering is a quantum mechanical process. This prohibits the separation of plasmonic antenna and Raman scatterer into two distinct subsystems as done in the EM model. Here we propose a quantum theory of PERS where the plasmon forms an integral part of the excitation. It explains the experimental PERS profile, predicts the appearance of defect-type phonons, and shows ways for manipulating PERS enhancement. Consider a plasmonic nanostructure in close proximity to a Raman scatterer such as a molecule or graphene, which will be referred to as the Raman probe (Fig. 3a). PERS is then described as one quantum mechanical process as depicted by Feynman diagrams in Fig. 3b-e.[25,36] It starts with the excitation of the LSPR by the incoming photon $E_1=\hbar\omega_1$ [step (1) in Fig. 3a and b]. The plasmon couples (2) via its near field to the electrons of the probe creating an electronic excitation.[38] The excited carriers emit (3) a phonon $\hbar\omega_{ph}$ and recombine (4) by coupling again to the LSPR. The plasmonic nanostructure radiates (5) the scattered photon $E_2=\hbar\omega_2$. Our key argument is that three other scattering processes will result in the same final state (phonon $\hbar\omega_{ph}$ excited and photon $\hbar\omega_2$ emitted). They are the Raman process without any coupling to the LSPR (Fig. 3c) and two processes where either incoming (Fig. 3d) or outgoing photon (Fig. 3e) couple to the plasmon. When calculating the PERS intensity we have to allow for interferences between the scattering channels. Mathematically this means that we have to sum the amplitudes resulting from Fig. 3b-e before squaring to calculate the intensity.

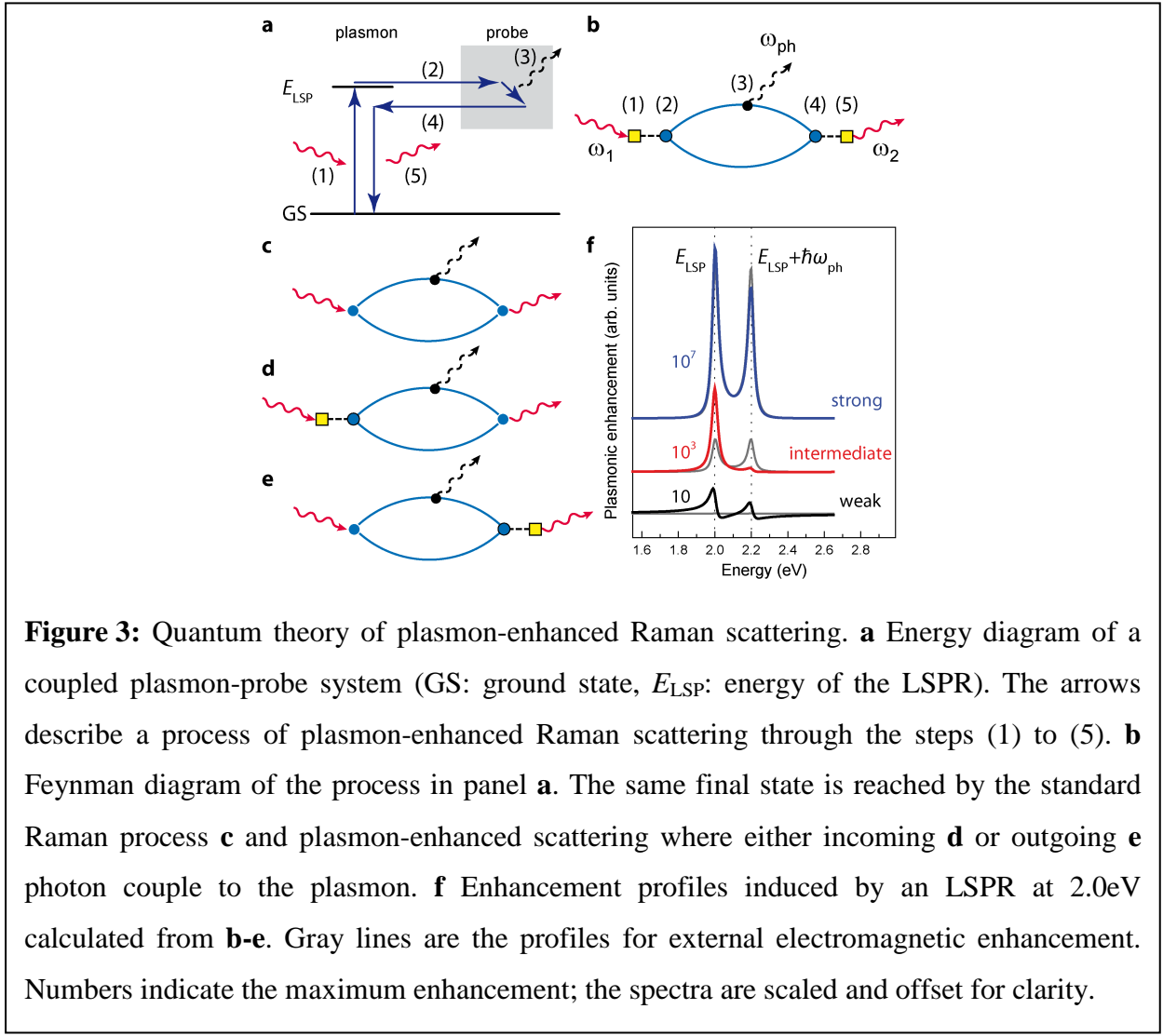

**Figure 3:** Quantum theory of plasmon-enhanced Raman scattering. **a** Energy diagram of a coupled plasmon-probe system (GS: ground state, $E_{LSP}$: energy of the LSPR). The arrows describe a process of plasmon-enhanced Raman scattering through the steps (1) to (5). **b** Feynman diagram of the process in panel **a**. The same final state is reached by the standard Raman process **c** and plasmon-enhanced scattering where either incoming **d** or outgoing **e** photon couple to the plasmon. **f** Enhancement profiles induced by an LSPR at 2.0eV calculated from **b-e**. Gray lines are the profiles for external electromagnetic enhancement. Numbers indicate the maximum enhancement; the spectra are scaled and offset for clarity.

We derive an expression for the PERS intensity $I_{ph}^{\text{PERS}} \propto \left|K_{ph}^{\text{PERS}}\right|^2$ and its dependence on excitation energy using the microscopic theory of Raman scattering.[36] We assume a single LSPR to simplify the treatment; the formalism can be expanded to multiple excitations by summing over intermediate states. We consider Stokes scattering by $q=0$ phonons (phonon emission during scattering, first-order Raman effect). The diagram in Fig. 3b translates into a scattering probability by the Fermi Golden Rule. It contributes the following term to the matrix element of PERS

$$K_{ph,b}^{\text{PERS}} = \sum_e \frac{M_{pt-pl}M_{pl-e}M_{e-ph}M_{e-pl}M_{pl-pt}}{(E_1-E_{LSP}-i\gamma_{LSP})(E_1-E_e-i\gamma_e)(E_2-E_e-i\gamma_e)(E_2-E_{LSP}-i\gamma_{LSP})}$$
$$= K_{ph}^{\text{Raman}} \frac{\tilde{M}^2}{(E_1-E_{LSP}-i\gamma_{LSP})(E_2-E_{LSP}-i\gamma_{LSP})}. \qquad (2)$$

$\tilde{M}$ is a measure of the plasmon-probe coupling; it is the ratio between the matrix elements for plasmon-mediated and direct excitation of an electron in graphene.

$$\tilde{M} = \frac{M_{pt-pl}M_{pl-e}}{M_{pt-e}} = \frac{M_{e-pl}M_{pl-pt}}{M_{e-pt}} \qquad (3)$$

$K_{ph}^{\text{Raman}}$ is the Raman cross section (constant in graphene)

$$K_{ph}^{\text{Raman}} = \sum_e \frac{M_{pt-e}M_{e-ph}M_{pt-e}}{(E_1-E_e-i\gamma_e)(E_2-E_e-i\gamma_e)}. \qquad (4)$$

$E_{LSP}$ is the energy of the LSPR and $2\gamma_{LSP}$ its width. The electronic state of the probe with energy $E_e$ and lifetime $\gamma_e$ is assumed to be the same before and after phonon emission ($q=0$ intraband scattering processes). All intermediate electronic states *e* need to be summed over when evaluating Eq. (2). The matrix elements describe photon-plasmon coupling $M_{pt-pl}$, photon-electron coupling $M_{pt-e}$, plasmon-electron coupling $M_{pl-e}$, and electron-phonon coupling $M_{e-ph}$. They are assumed to be independent of wavevector and identical for excitation and recombination. We restrict the calculation to the most resonant time order.[36,39]

The other Feynman diagrams in Fig. 3 yield: Figure 3c (Raman term)

$$K_{ph,c}^{PERS} = K_{ph}^{Raman}, \qquad (5)$$

Fig. 3d (coupling only the incoming photon to the LSPR)

$$K_{ph,d}^{PERS} = K_{ph}^{Raman} \frac{\tilde{M}}{(E_1 - E_{LSP} - i\gamma_{LSP})}, \qquad (6)$$

and Fig. 3e (coupling only the outgoing photon to the LSPR)

$$K_{ph,e}^{PERS} = K_{ph}^{Raman} \frac{\tilde{M}}{(E_2 - E_{LSP} - i\gamma_{LSP})}. \qquad (7)$$

The total cross section for plasmon enhanced Raman scattering is obtained by summing the four Feynman diagrams

$$\begin{aligned}K_{ph}^{PERS} &= K_{ph,b}^{PERS} + K_{ph,c}^{PERS} + K_{ph,d}^{PERS} + K_{ph,e}^{PERS} \\ &= K_{ph}^{Raman}(1 + \frac{\tilde{M}}{E_1 - E_{LSP} - i\gamma_{LSP}} + \frac{\tilde{M}}{E_2 - E_{LSP} - i\gamma_{LSP}} + \frac{\tilde{M}^2}{(E_1 - E_{LSP} - i\gamma_{LSP})(E_2 - E_{LSP} - i\gamma_{LSP})}),\end{aligned}$$
(8)

The absolute square of Eq.(8) is proportional to the PERS intensity. For strong plasmon-probe coupling $\tilde{M}$ the term in Fig. 3b dominates plasmon-enhanced Raman scattering. The enhancement profile predicted from the quantum theory of PERS in Fig. 3f (blue line) is close to the profile from electromagnetic enhancement (gray). In the intermediate coupling regime (red line in Fig. 3f) the processes in Fig. 3b, d, and e have similar probability. The incoming resonance at $E_{LSP}$ increases in intensity due to a constructive quantum interference between the scattering channels in Fig. 3b and d, whereas the outgoing resonance at $E_{LSP}+\hbar\omega_{ph}$ is almost completely absent due to destructive interference between Fig. 3b and e. In the weak coupling regime the terms in Fig. 3d and e dominate, resulting in detectable resonances (black line in

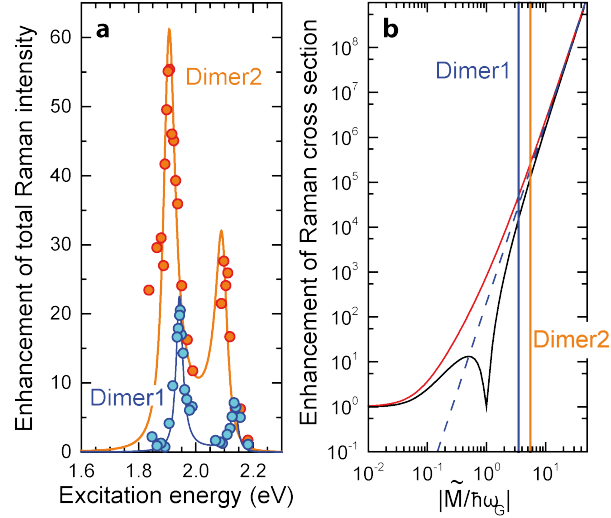

**Figure 4: a** PERS profile with the background subtracted as measured on Dimer1 and Dimer2. **b** Plasmonic enhancement calculated from Eq.(8) ($\hbar\omega_G$=0.2eV, $E_{LSP}$=2eV, $\gamma_{LSP}$=0.02eV). The red line corresponds to the maximum of the outgoing (incoming) resonance for positive (negative) $\tilde{M}$; the black line to the incoming (outgoing) resonance for positive (negative) $\tilde{M}$. Vanishing enhancement at $|\tilde{M}/\hbar\omega_G|=1$ is a universal prediction. The dashed line is for electromagnetic enhancement; the vertical lines mark the coupling parameters measured on Dimer1 and Dimer2.

Fig. 3f) when the predicted EM enhancement vanishes (gray). We fit the experimentally observed plasmon-enhanced intensity by $I_G^{PERS} \propto |K_G^{PERS}|^2$ with the cross section given in Eq.(8). From the PERS enhancement profile of Dimer1 in Fig. 4a we obtain $E_{LSP}$=1.942eV and $\gamma_{LSP}$=14meV. The energy agrees with the maximum in elastic scattering measured with dark-field spectroscopy (Supplementary Fig. S3). Dimer2 had a slightly lower $E_{LSP}$=1.905 eV, a larger $\gamma_{LSP}$=25meV, and stronger enhancement (Fig. 4a).

Equation (8) predicts interference between the scattering pathways depending on the magnitude and sign of $\tilde{M}$. The PERS profiles yielded $\tilde{M}_1$=−(0.7±0.2) eV for Dimer1 and $\tilde{M}_2$=−(1.1±0.1) eV for Dimer2 (Fig. 4a). We now further discuss the PERS profiles and the polarization dependence of the enhancement. This analysis will allow instructive cross checks for our proposed theory. The coupling term $\tilde{M}$ and the resonance width $\gamma_{LSP}$ for a given dimer allow predicting the enhancement of the Raman cross section. From Fig. 2b we obtain $(5\pm1)\cdot10^4$ enhancement at the incoming resonance of Dimer1 and $(3\pm1)\cdot10^5$ for Dimer2, which is reasonably close to the geometry-based estimate in Fig. 1b. The higher coupling term of Dimer2, furthermore, implies a smaller ratio between the intensity at incoming and outgoing resonance, Fig.4a. Indeed $I_{in}/I_{out}=3$ for Dimer1, but only 2 for Dimer2. We also compare the enhancement ratio between Dimer2 and Dimer1 $(6\pm2)$ calculated from the coupling parameters with the measured enhancement ratio of $(8\pm1)$, Fig. 4a after scaling the peak heights to identical FWHM. Although the uncertainties in our analysis are large, we obtain a coherent picture when comparing our model for plasmon-enhanced Raman scattering to the measured data.

The difference between the coupling parameters $\tilde{M}_1$ and $\tilde{M}_2$ is caused by a varying graphene-plasmon distance from Dimer1 to Dimer2. The graphene is almost flat on top of Dimer1, but pulled into the near-field cavity of Dimer2 as demonstrated by AFM topography images in Supplementary Fig. S2.[17,19] We verified this interpretation by determining the strain in graphene on top of the dimers from the Raman frequency shift of the *G* line, i.e., the part that is enhanced by the plasmonic near field. We obtain 2% strain on Dimer2, but only 0.4% on Dimer1 (Supplementary Information). The higher strain is in excellent agreement with the AFM observations. The graphene covering Dimer2 is closer to the hotspot resulting in a stronger plasmonic enhancement expressed through the larger $\tilde{M}$.

SERS and TERS emerge from the quantum theory of PERS as the limit of strong and weak coupling. Single-molecule SERS requires an enhancement on the order of $10^7$-$10^9$ (Refs 25 and 29). According to Fig. 4b this corresponds to coupling parameters $|\tilde{M}/\hbar\omega_{ph}|=10$-$50$, which is only one order of magnitude higher than the coupling observed in the nanodimer-graphene system ($|\tilde{M}/\hbar\omega_{ph}|=5$). The graphene in our configuration is comparatively far away from the plasmonic hotspot. A placement of, e.g, a nanotube or a molecule in the cavity near field would provide an even higher enhancement. A rational design of SERS hotspots with single-molecule sensitivity appears within reach after further refinement of the plasmonic nanostructure. The range of TERS enhancement resides around $|\tilde{M}/\hbar\omega_{ph}|\approx 1$ (Fig. 4b). The interference between the scattering pathways strongly increases one of the channels in this coupling regime making TERS a sensitive near-field probe. The second resonance channel, however, is suppressed by destructive interference. Indeed a missing incoming resonance was accompanied by a strong outgoing resonances in a molecular TERS experiments.[10]

The quantum theory of PERS has striking consequences on the polarization dependence and the defect-induced modes of graphene. Experiments with polarized light will allow an independent determination of $\widetilde{M}$. We will also show that the *D* and *D'* mode (Fig. 1b) are expected to appear for high-quality plasmonic resonances. Coupling to the LSPR of the nanocavity requires ∥ polarization along the dimer axis, because the hotspot is strongly polarized.[17,23] So far, we have worked experimentally in the (∥,∥) scattering configuration, i.e., both photons were polarized along the dimer. In deriving Eq.(8) we likewise assumed that both incoming and outgoing photon are allowed to couple to the LSPR, i.e., again (∥,∥) configuration. Coupling to the LSPR of the nanocavity is forbidden when both incoming and outgoing light are polarized perpendicular to the dimer, (⊥,⊥) configuration. Heeg *et al.*[17] showed that the (⊥,⊥) configuration results in weak plasmonic enhancement that is delocalized over the two gold discs. This behavior is confirmed at the incoming and outgoing resonance of Dimer1. When switching from (∥,∥) into (⊥,⊥) configuration the *G* line intensity drops from maximum enhancement to an intensity that equals the delocalized enhancement (background in Fig. 2a, Supplementary Fig. S8 and discussion in the text). The *G* peak of graphene is independent of polarization,[22,36] any change in Raman intensity reflects a change in plasmonic enhancement.

Even more intriguing are the Raman intensities under crossed polarization of the incoming and scattered light. Crossed polarization prohibits only certain scattering pathway in Fig. 3b, while others remain allowed. The term in Fig.3b and Eq.(2) is prohibited in both (∥,⊥) and (⊥,∥) configuration. Additionally, the term in Fig. 3d/Eq.(6) is prohibited in the (⊥,∥) configuration and Fig. 3e/Eq.(7) in the (∥,⊥) configuration. This means that selected scattering pathways or certain plasmonic states can be addressed selectively using polarized Raman spectroscopy. Specifically, for an outgoing resonance excited at a laser energy $E_1=E_{\text{LSP}}+\hbar\omega_G$ we find and intensity ratio [Eq.(8)]

$$\frac{I_G^{\text{PERS}}(\parallel,\parallel)}{I_G^{\text{PERS}}(\perp,\parallel)} \approx \left|1+\widetilde{M}/\hbar\omega_G\right|^2. \qquad (2)$$

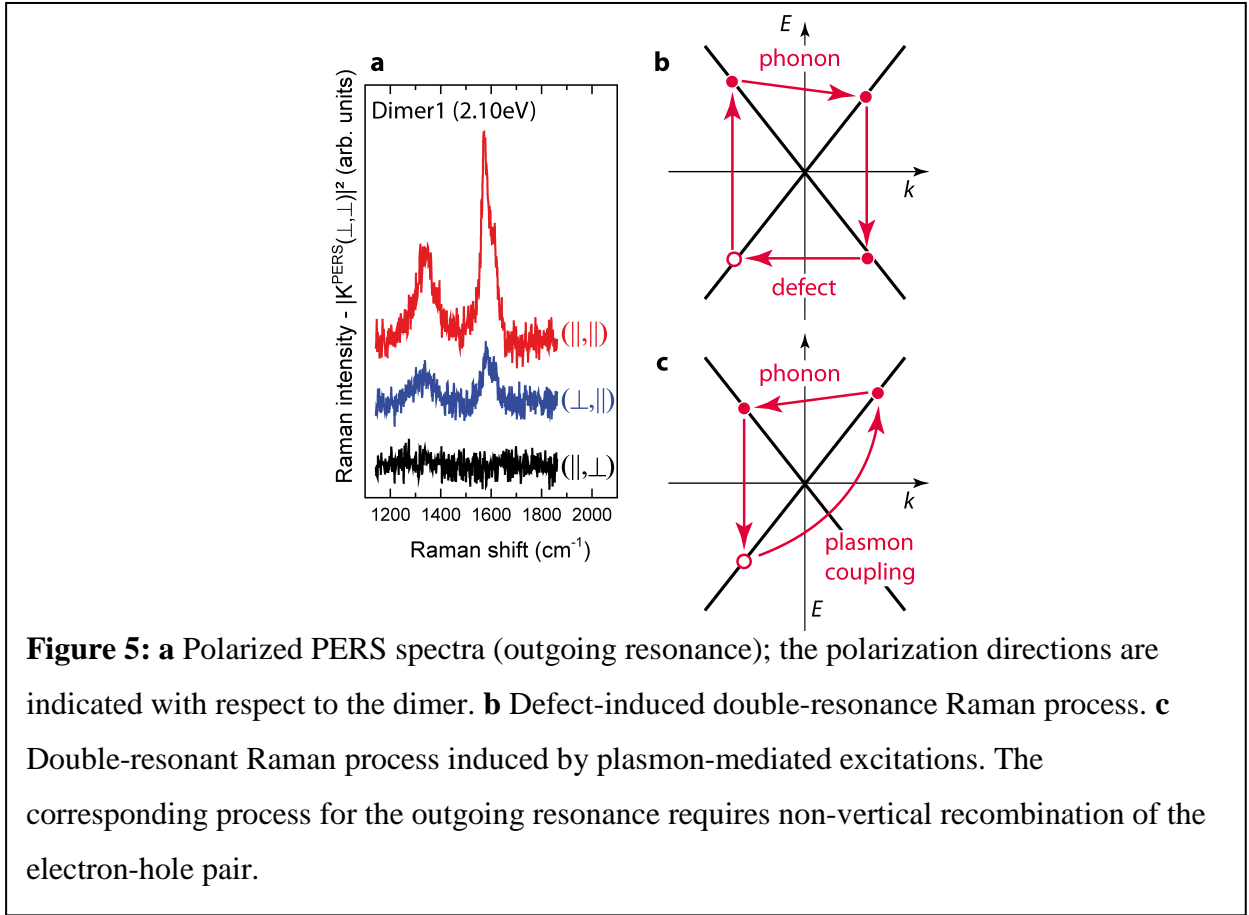

**Figure 5: a** Polarized PERS spectra (outgoing resonance); the polarization directions are indicated with respect to the dimer. **b** Defect-induced double-resonance Raman process. **c** Double-resonant Raman process induced by plasmon-mediated excitations. The corresponding process for the outgoing resonance requires non-vertical recombination of the electron-hole pair.

The polarized spectra thus allow to independently determine the plasmonic coupling parameter $\tilde{M}$. In Fig. 5a we show the Raman spectra of Dimer1 excited at the outgoing resonance after subtracting the delocalized background enhancement, compare Supplementary Information. The ratio $I_G^{\text{PERS}}(\|,\|)/I_G^{\text{PERS}}(\perp,\|) = 4$ yields a coupling parameter $\tilde{M}_1 = -0.6$ eV in excellent agreement with the fit of the PERS profile ($\tilde{M}_1 = -0.7$ eV). The polarization dependence further verifies our proposed theory of plasmon-enhanced light scattering. Note also the vanishing intensity in $(\|,\perp)$ configuration, Fig. 5a. It confirms that the plasmonic resonance is much narrower than the phonon energy as already observed in the enhancement profiles.

Finally, we explain the appearance of the *D* and *D'* peaks within the quantum theory of PERS. The origin of the defect lines in standard Raman scattering is a double-resonant Raman process involving defect-scattering as one step.[30] The scattering process (fourth-order in perturbation theory) is depicted in Fig. 5b. The strongly localized nanocavity plasmon excites non-vertical intraband excitations (Fig. 5c).[40] This is followed by phonon emission and the recombination of the excited electron-hole pair. The plasmon-mediated excitation thus activates a double-resonant Raman process without requiring any defect for wavevector conservation.[30] The dominant phonon frequencies in the experimental spectra are selected by the double-resonance condition, see Ref. 30.

The intensity ratio of the *D* and *G* peak $I_D/I_G$ has a theoretical limit of ten if the *D* mode gets activated for identical photon-electron coupling in the *G* and *D* process, perfect elastic scattering, and no loss in scattering material.[41] In view of the different plasmon-electron coupling in *G* and *D* mode PERS, the smaller experimental value of three is reasonable (Fig. 1b). The intensity of the *D* line scales with the characteristic localization length $I_D \sim 1/L_D^2$.[41,42] Spectra recorded in the ($\perp,\perp$) configuration are enhanced by the delocalized near field over the entire dimer,[17,35] which explains the absence of the defect modes in perpendicular polarization and confirms the quantum nature of plasmon-enhanced Raman scattering. The activation of the *D* and *D'* modes are fingerprints for the near-field localization by a cavity. The mode can be used in screening for high-quality plasmonic devices.

In conclusion, we measured plasmon-enhanced Raman scattering in graphene on an individual hotspot. The strong enhancement by a rationally designed gold nanodimer demonstrated the power of lithographically fabricated plasmonic hotspots. In wavelength-dependent measurements of PERS we observed narrow pairs of incoming and outgoing resonances induced by the localized surface plasmon. We proposed a quantum theory of PERS as a powerful formalism to model and predict plasmonic enhancement. For a Raman probe close to a plasmonic hotspot the excitation of phonons occurs through four competing scattering channels; their interference results, e.g., in suppressed plasmonic resonance. The strongly localized hotspot of a plasmonic cavity activates scattering by defect-type phonons in graphene. They can be used to characterize the localization and quality factor of a PERS hotspot. Our work unifies the existing models for SERS and TERS, which emerge as the limiting cases for strong and weak coupling. We highlighted the intermediate regime of plasmonic coupling and its previously unknown quantum interferences. This regime is of particular technological importance because it governs the plasmonic enhancement of photocurrent and photodetection in one- and two-dimensional nanostructures.[14,18]

**Acknowledgement:** This work was supported by the ERC (Grant 210642) and the NanoScale Focus Area. SR, PK, SH, and CL acknowledge the Deutsche Forschungsgemeinschaft (SFB658 and Nanospec). AV, SH, AO, and NC acknowledge the Engineering and Physical Sciences Research Council EP/G035954/1 and EP/K009451/1.

**Author contribution:** The quantum theory of plasmon-enhanced Raman scattering was suggested by SR and worked out together with SH and NSM. SH, SR, and AV designed the nanodimer-graphene system that was fabricated by AO, NC, and AV. SH carried out Raman experiments with fixed excitation energy and quantified the enhancement. PK, CL, and SW recorded the PERS spectra presented in this study and analyzed them together with NSM and SR. The manuscript was written by SR, discussed and revised by all co-authors.

**Supporting Information Available:** Supporting Information and Figures that discuss the characterization of the nanodimers, the polarization-dependent Raman measurements, and strain analysis. This material is available free of charge via the Internet at http://pubs.acs.org.

**References**

[1] L. Novotny, B. Hecht, *Principles of Nano-Optics*, Cambridge (2012).

[2] S. A. Maier, *Plasmonics: Fundamentals and Application* (New York, Springer, 2007).

[3] M. Fleischmann, P. J. Hendra, A. J. McQuillan, Raman spectra of pyridine adsorbed at a silver electrode, Chem. Phys. Lett. **26**, 2 (1974).

[4] K. Kneipp, Y. Wang, H. Kneipp, I. Itzkan, R. R. Dasari, M. S. Feld, Population Pumping of Excited Vibrational States by Spontaneous Surface-Enhanced Raman Scattering, Phys. Rev. Lett. **76,** 2444 (1996).

[5] T. A. Laurence, G. B. Braun, N. O. Reich, M. Moskovits, Robust SERS Enhancement Factor Statistics Using Rotational Correlation Spectroscopy, Nano Lett. **12**, 2912 (2012).

[6] E. C. Le Ru, E. Blackie, M. Meyer, P. G. Etchegoin, Surface Enhanced Raman Scattering Enhancement Factors: A Comprehensive Study, J. Phys. Chem. C **111**, 13794 (2007).

[7] J. A. Dieringer et al., Surface enhanced Raman spectroscopy: new materials, concepts, characterization tools, and applications, Far. Disc. **132**, 9 (2006).

[8] P. G. Etchegoin, C. E. Le Ru, Single-Molecule Surface-Enhanced Raman Spectroscopy, Annu. Rev. Phys. Chem. **63**, 65 (2012).

[9] B. Pettinger, B. Ren, G. Picardi, R. Schuster, G. Ertl, Nanoscale Probing of Adsorbed Species by Tip-Enhanced Raman Spectroscopy, Phys. Rev. Lett. **92**, 096101 (2004).

[10] R. Zhang, *et al.*, Chemical mapping of a single molecule by plasmon-enhanced Raman scattering, Nature **498**, 82 (2013).

[11] *Tip Enhancement* ed. S. Kawate and V. M. Skalaev (Elsevier, 2007).

[12] V. Giannini, et al., Controlling Light Localization and Light–Matter Interactions with Nanoplasmonics, Small 6, 2498 (2010);

[13] Collection Surface Enhanced Raman Spectroscopy, Chem. Soc. Rev. **5**, 873-1076 (2008).

[14] M. L. Brongersma, N. J. Halas, P. Nordlander, Plasmon-induced hot carrier science and technology, Nat. Nanotech. **10**, 25 (2015).

[15] Z. Fang et al., Graphene-antenna sandwich photodetector, Nano Lett. 12, 3808 (2012).

[16] F. Schedin, et al., Surface-Enhanced Raman Spectroscopy of Graphene, ACS Nano **4**, 5617 (2010).

[17] S. Heeg, *et al.*, Polarized Plasmonic Enhancement by Au Nanostructures Probed through Raman Scattering of Suspended Graphene, Nano Lett. **13**, 301 (2013).

[18] V. G. Kravets *et al.*, Surface Hydrogenation and Optics of a Graphene Sheet Transferred onto a Plasmonic Nanoarray, J. Chem. Phys. C **116**, 3882 (2012).

[19] S. Heeg *et al.*, Strained graphene as a local probe for plasmon-enhanced Raman scattering by gold nanostructures, Phys. Stat. Sol. (RRL) **7**, 1067 (2013).

[20] S. Reich, C. Thomsen, Raman spectroscopy of graphite, Phil. Trans. A **362**, 2271 (2004).

[21] A. C. Ferrari, D. M. Basko, Raman spectroscopy as a versatile tool for studying the properties of graphene, Nat. Nanotech. **8**, 235 (2013)

[22] M. S. Dresselhaus, A. Jorio, M. Hofmann, G. Dresselhaus, R. Saito, Perspectives on Carbon Nanotubes and Graphene Raman Spectroscopy, Nano Lett. **10**, 751 (2010).

[23] S. Heeg *et al.*, Plasmon-Enhanced Raman Scattering by Carbon Nanotubes Optically Coupled with Near-Field Cavities, Nano Lett. **14**, 1762 (2014).

[24] Heeg *et al.*, Plasmon-enhanced Raman scattering by suspended carbon nanotubes, phys. stat. sol. rrl **8**, 785 (2014).

[25] M. Moskovits, Persistent misconceptions regarding SERS, Phys. Chem. Chem. Phys. 15, 5301 (2013).

[26] P. Alonso-González, *et al.*, Resolving the electromagnetic mechanism of surface-enhanced light scattering at single hot spots, Nature Comm. **3**, 684 (2012).

[27] D. Yoon et al., Interference effect on Raman spectrum of graphene on $SiO_2$/Si, Phys. Rev. B **80**, 125422 (2009).

[28] J. A. Dieringer *et al.*, Surface-Enhanced Raman Excitation Spectroscopy of a Single Rhodamine 6G Molecule, JACS **131**, 849 (2009).

[29] Y. Fang, N.-H. Seong, D. D. Dlott, Measurement of the Distribution of Site Enhancements in Surface-Enhanced Raman Scattering, Science **321**, 388 (2008).

[30] C. Thomsen, S. Reich, Double Resonant Raman Scattering in Graphite, Phys. Rev. Lett. **85**, 5214 (2000).

[31] M. D. Doherty, A. Murphy, J. McPhillips, R. J. Pollard, P. Dawson, Wavelength Dependence of Raman Enhancement from Gold Nanorod Arrays: Quantitative Experiment and Modeling of a Hot Spot Dominated System, J. Phys. Chem. **114**, 19913 (2010).

[32] A. D. McFarland, M. A. Young, J. A. Dieringer, R. P. Van Duyne, Wavelength-Scanned Surface-Enhanced Raman Excitation Spectroscopy, J Phys Chem B **109**, 11279 (2005).

[33] E. C. Le Ru, C. Galloway, P. G. Etchegoin, On the connection between optical absorption/extinction and SERS enhancements, Phys. Chem. Chem. Phys. **8**, 3083 (2006).

[34] W. Zhu, K. B. Crozier, Quantum mechanical limit to plasmonic enhancement as observed by surface-enhanced Raman scattering, Nat. Comm. DOI: 10.1038/ncomms6228 (2015).

[35] I. Khan *et al.*, From Micro to Nano: Analysis of Surface-Enhanced Resonance Raman Spectroscopy Active Sites via Multiscale Correlations, Anal. Chem. **78**, 224 (2006).

[36] M. Cardona, *Resonance Phenomena*, in *Light Scattering in Solids II* (Springer, 1982, New York).

[37] E. LeRu, P. Etchegoin, Rigorous justification of the $|E|^4$ enhancement factor in Surface Enhanced Raman Spectroscopy, Chem. Phys. Lett. **423**, 63 (2006).

[38] T.-K. Lee, J. L. Birman, Molecule adsorbed on plane metal surface: Coupled system eigenstates, Phys. Rev. B **22**, 5953 (1980).

[39] These approximations are commonly used in Raman scattering. Neglecting the non-resonance time orders has negligible effect on calculating resonant Raman intensities (error $10^{-8}$ or less). The error introduced by setting the matrix elements constant and independent of wavevector is not known. It may affect the strength of the coupling and theh magnitude of the coupling parameter. However, relative magnitudes and the form of the enhancement profiles remain valid.

[40] F. H. L. Koppens, D. E. Chang, F. J. García de Abajo, Graphene Plasmonics: A Platform for Strong Light–Matter Interactions, Nano Lett. **11**, 3370 (2011).

[41] L. G. Cancado *et al.,* Quantifying Defects in Graphene via Raman Spectroscopy at Different Excitation Energies, Nano Lett. **11**, 3190 (2011).

[42] J. F. Rodriguez-Nieva, E. B. Barros, R. Saito, M. S. Dresselhaus, Disorder-induced double resonant Raman process in graphene, Phys. Rev. B **90**, 235410 (2014).

**Supplementary Information**

See separate file

# Supplementary Information - Quantum Nature of Plasmon-Enhanced Raman Scattering

Patryk Kusch, Sebastian Heeg, Christian Lehmann, Niclas S. Müller, Sören Wasserroth, Antonios Oikonomou, Nicholas Clark, Aravind Vijayaraghavan, Stephanie Reich

## 1. Dimer geometry and imaging by SEM and AFM

Gold nanodimers were produced by electron beam lithography as described in the main paper. We characterized the dimers by SEM, AFM, Raman imaging, and dark field spectroscopy before focusing on single nanodimers for wavelength-dependent Raman measurements. In Supplementary Fig. S1 we show SEM pictures taken on 16 nanodimers demonstrating highly uniform geometries. The pictures were taken on a Zeiss Ultra Plus SEM with 10kV acceleration voltage. The gold nanodisc diameters vary from 100 to 115nm with a mean diameter (110±5) nm. The dimer gap ranged from 18 to 25 nm with a mean of (21±4) nm. In some panels in Fig. S1 the two discs appear to be connected; however, this is not gold, but resist (PMMA) that was cross-linked due to proximity effects during electron beam exposure.

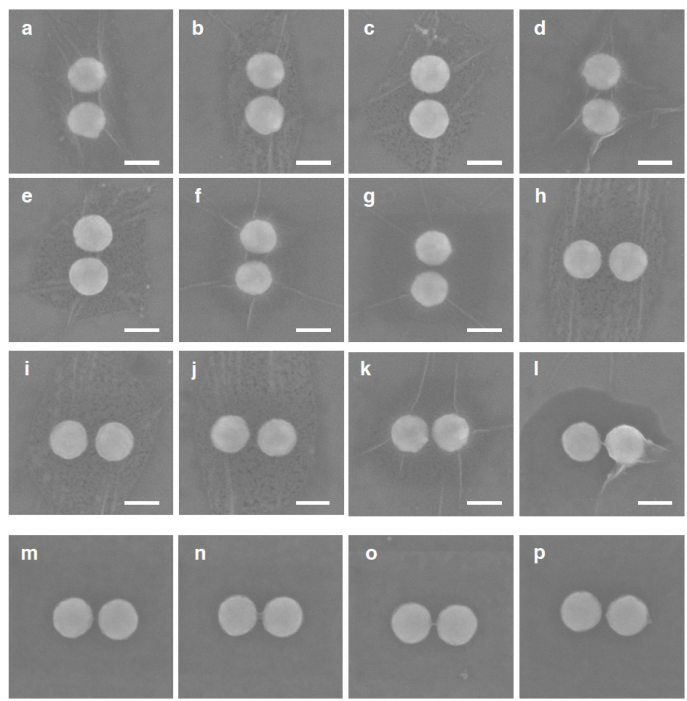

**Figure S1:** Gold nanodimers with graphene-cover layer (a-l) and without it (m-p). Compared to the AFM, Fig S2, the graphene configuration on and around the dimers is modifying by the SEM measurements themselves due to the exposure to vacuum conditions and beam induced damage of the graphene membrane. The scale bar is 100nm in all panels.

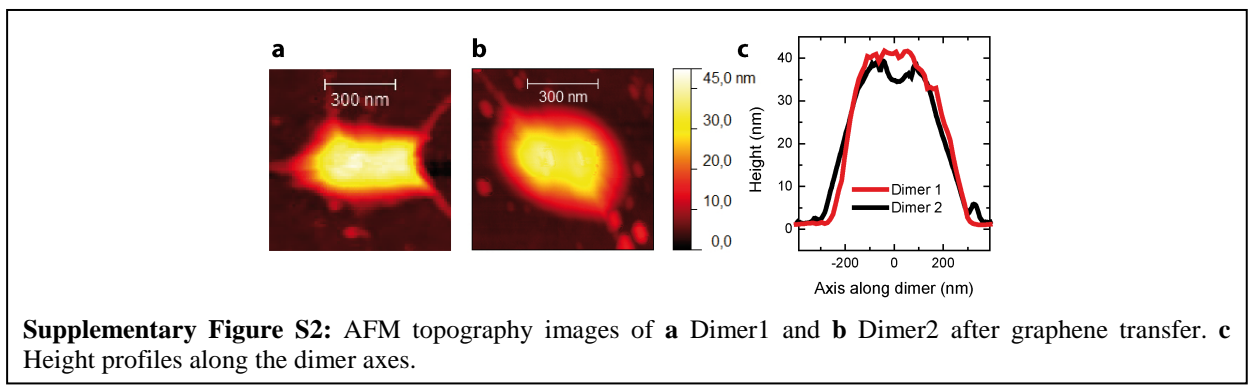

**Supplementary Figure S2:** AFM topography images of **a** Dimer1 and **b** Dimer2 after graphene transfer. **c** Height profiles along the dimer axes.

The SEM pictures of nanodimers covered with graphene Fig.S1a-f and l-p show the strain and wrinkles created in graphene by the nanodimers after the polymer-support transfer. We note that the SEM pictures were taken under vacuum conditions; under ambient conditions the graphene membrane adheres closely to the gold nanodiscs as verified by AFM.

Atomic force microscopy of Dimer1 and Dimer2 was performed after the graphene transfer using a Park System XE 150 AFM in tapping mode. Similar images were obtained in Ref. 1, where we compared them to the distinct topography of the pristine nanodimer. The topography after transfer resembles a double-dot structure covered by a membrane, Fig. S2a and b. Height profiles parallel to the axes were cut through the center of the dimers, Fig. S2c. They indicate that the graphene is pulled more strongly into the void of the nanocavity in Dimer2, see black trace in Fig. S2c. In Dimer1 the height profile appears flat on top of the nanodimer. A detailed discussion of the topography and strain configuration of graphene covering plasmonic dimers is given in Refs. 1 and 2. The references also contain more details on nanodimer fabrication and characterization, e.g. AFM images before and after graphene transfer and different geometries of the plasmonic hotspots (monomers, trimers).

## 2. Light scattering: Raman scans and dark-field spectroscopy

The plasmonic nanodimers were further studied by elastic scatting (dark field spectroscopy) and Raman scans over dimer fields. Elastic scattering of light by the gold nanodimers was excited with a white light lamp (50W power). It was focused onto the sample and collected with a dark-field 100x objective and a fiber. The light was dispersed in a single-grating spectrometer and detected with a CCD. The dark-field spectra of Dimer1 and Dimer2 are shown in Supplementary Fig. S3 together with the PERS profile measured by tunable Raman scattering. The maxima in dark-field spectroscopy of these and other dimers were between 1.8 and 2.0 eV with a total FWHM 200 meV.

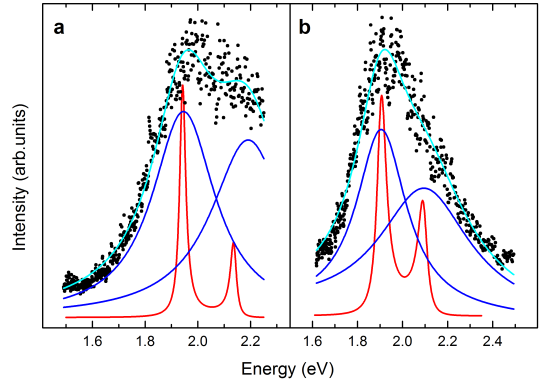

**Supplementary Figure S3:** Dark-field spectra. **a** Dimer1 and **b** Dimer2. Black dots are the measured data that were fit (cyan) to two Lorentzians (blue line). For comparison the PERS profiles are shown (red, see Fig.4a).

For a first characterization of plasmon-induced Raman enhancement we collected Raman spectra at a fixed laser excitation energy on rows of dimers. These experiments were performed on single-grating Horiba XploRa and WITEC Raman spectrometers. They are equipped with nanopositioning scanning stages. A key characteristic of these systems is their extremely high throughput and sensitivity as well as small diameters of the laser focus, which makes them ideal for scanning dimer fields. On the other hand, they are limited by construction to excitations at fixed laser energies (in the red the XPlora has a 638nm diode laser and the WITEC a 633nm HeNe laser). Also the frequency resolution of the spectrometers is poor compared to triple-grating systems. The much more cumbersome wavelength- and polarization-dependent studies, therefore, had to be performed on the fully tunable Horiba T64000 spectrometer.

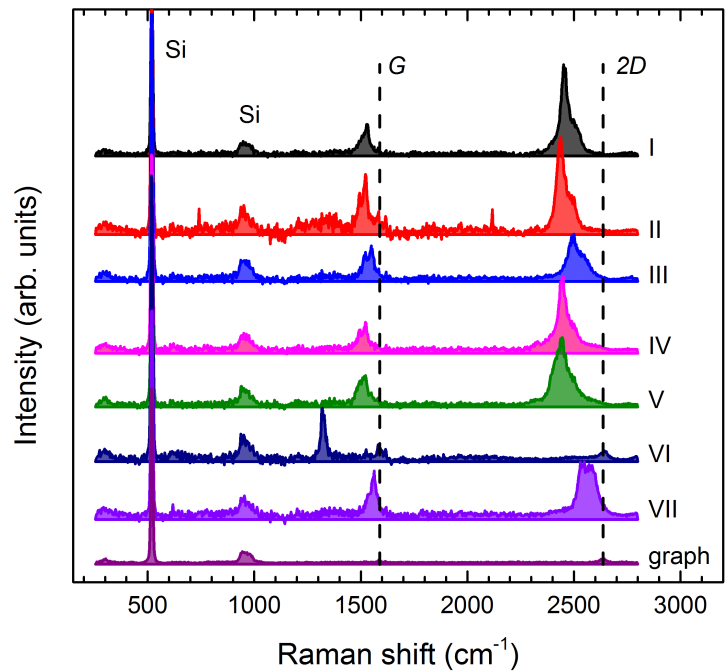

**Supplementary Figure S4:** Raman spectra obtained on seven neighboring hotspots (I – VII) in a dimer line. The lowest trace (graph) shows the spectrum of graphene as is measured between the dimers and far away from the plasmonic nanostructures. Six of the seven nanodimers induced plasmonic enhancement at 633nm as seen by the increasing scattering intensity compared to pristine graphene. The distance between the dimers was 10μm. The dashed vertical lines mark the positions of the G and 2D line observed in graphene. The label Si stands for the first and second-order Raman scattering by silicon. The spectra are presented as measured except for subtracting a broad luminescence background. Note the nearly-constant intensity of the Si second-order line (1000cm$^{-1}$) that serves as an internal reference in the as-measured spectra.

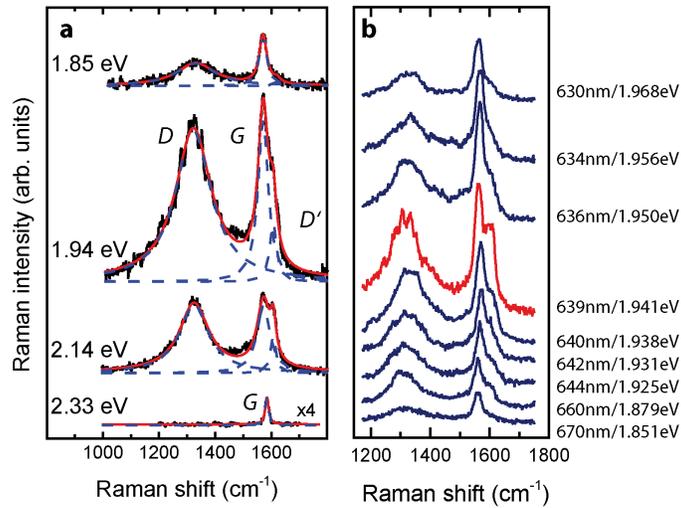

**Supplementary Figure S5: a** Exemplary fit of four Raman spectra measured on Dimer1 at different frequencies (labels). **b** Same data as shown in Fig.1d labeled with their excitation wavelength and energy.

Supplementary Figure S4 shows a typical scan where the laser spot was focused on seven consecutive dimers labeled I-VII. In the lowest trace we show the graphene spectrum that is observed between the nanodimers and away from the dimer line. The plasmon-induced enhancement of the Raman cross section leads to a strong increase in the scattering intensity of the $G$ and $2D$ peak of graphene in the dimers I to V and VII. The average increase in the total peak area is (36±6) for both lines (the values vary between 25 and 45 increase in total intensity). We note that these numbers must not be compared with the enhancement reported in the main part of the paper, because the focus size and the graphene reference are different from the main text. An increase in intensity in Fig. S4 is accompanied by a red shift of the phonon frequencies due to strain. Trace VI is the only spectrum without clear signatures of strong enhancement. Its phonon frequencies are identical with the graphene reference and there is only a very small increase in the scattering intensity, Fig.S4.

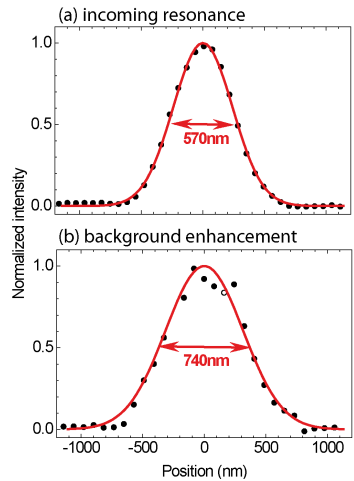

**Supplementary Figure S6:** Normalized intensity over the position of the laser spot with respect to the dimer center. **a** The laser energy matches the incoming resonance. The width of the scan profile is identical to the width of the laser focus. **b** Excitation away from the narrow resonance, but with a strong background enhancement. The width increased because the enhancement is now delocalized over both gold discs. See Ref. 1 for an in-depth discussion and similar measurements as a function of polarization.

Lithographic gold dimers very predictably increased Raman scattering by graphene. The dimer geometry of 110nm disc diameter and 20nm gap on a 300 nm $SiO_2$/Si substrate leads to plasmon-induced resonances close to 2eV. We conclude this from the maxima in the dark-field spectra as well as the strong increase in the Raman scattering intensity at 633 and 638nm.

Three dimers were selected for wavelength-dependent Raman measurements with fully tunable excitation. Two of the data sets are discussed in the main paper; a third resonance profile will be the topic of a future publication. To supplement the data analysis presented in Fig. 1 and 2 we show three exemplary fits of the measured spectra in Supplementary Fig. S5a. Note that there is no scattering at the frequency of the strained phonons for green excitation (2.33 eV). Supplementary Fig. S5b contains the wavelength-dependent measurements of Fig. 1d detailing all laser excitation energies used for the experiment.

We used a line scan to measure the width of the enhancement profile in real space when scanning over a dimer. Supplementary Fig. S6a shows a scan across the dimer when exciting in resonance with the highly localized nanocavity hotspot. When fitted with a Gaussian is has a FWHM= 570nm, which is identical to the diameter of the laser focus. Supplementary Fig. S6b presents a scan in resonance with the background enhancement. The width increased by 170 nm, which points towards scattering by two discs. See main paper for discussion.

**3. Plasmonic enhancement profiles from molecules**

The narrow FWHM of the plasmon-induced resonance as observed on graphene was an unexpected feature in plasmon-enhanced Raman scattering. In this section we compare the resonance profiles induced in graphene by a gold nanodimer to measurements by Dieringer *et al.*[3] and Zhu and Crozier[4] using molecular systems. Supplementary Fig. S6 compares the resonance profile measured on Dimer1 with the data reported by Dieringer *et al.*[3] on Rhodamin 6G on silver surfaces, Fig. S6b, and Zhu and Crozier[4] on thiophenol self-assembled monolayers on dimers of gold nanoparticles, Fig. S6c and d. The black dots are digitized data obtained from the references. The red lines are fits with Eq.(8). We see excellent agreement with the overall shape and width of the three experiments. The PERS enhancement profiles obtained on molecules are well described by our theory yielding a FWHM between 50 and 80 meV (see labels), which is comparable to our results (28 and 50 meV). Alternatively the data can also be fit by Lorentzians, which will result in almost identical FWHM. The similarity in the experimental profiles is truly remarkable, given the very different systems (plasmonic hotspot, Raman probe, see figure caption) from which these data were obtained. It clearly points toward a common fundamental mechanism in all three experiments.

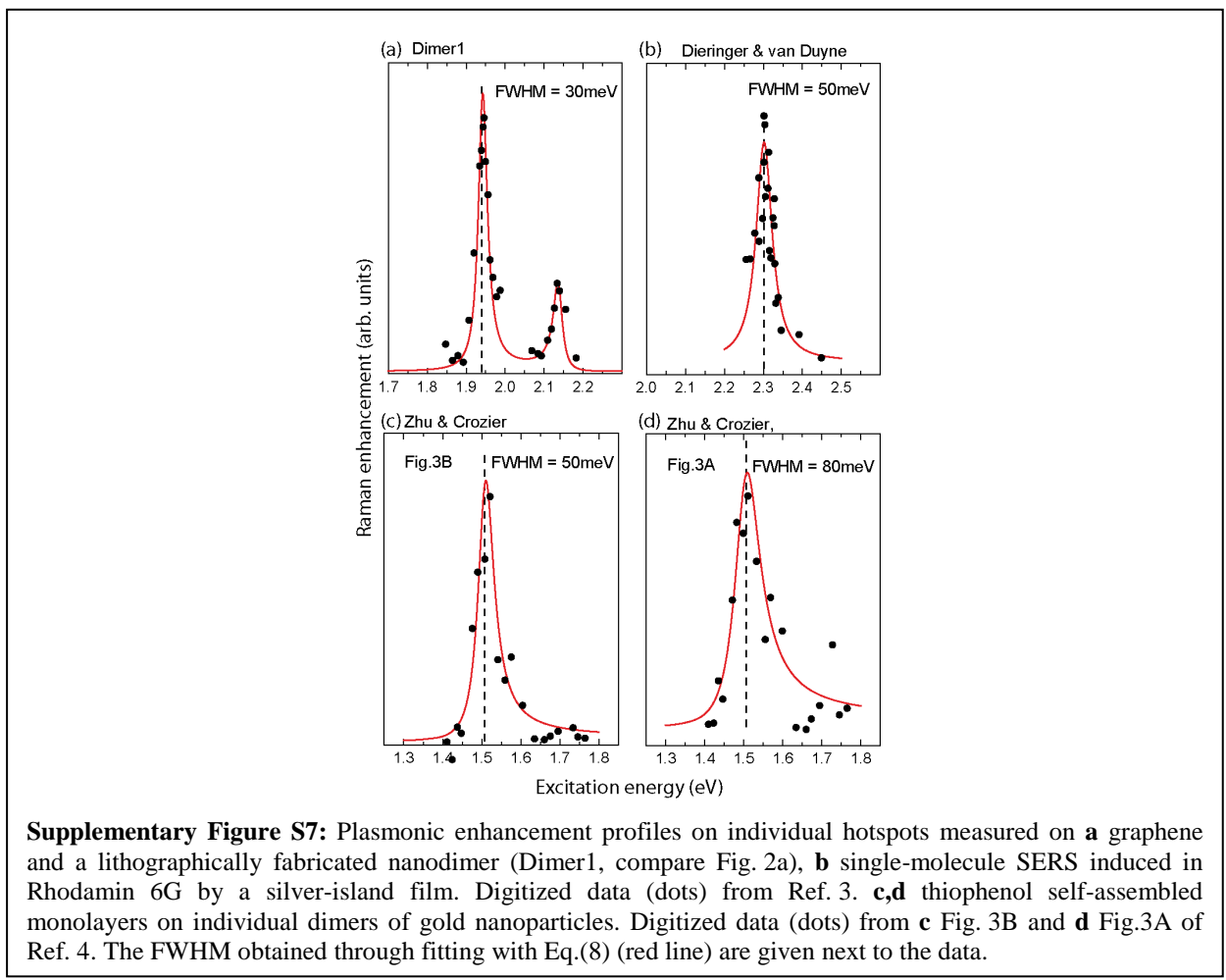

**Supplementary Figure S7:** Plasmonic enhancement profiles on individual hotspots measured on **a** graphene and a lithographically fabricated nanodimer (Dimer1, compare Fig. 2a), **b** single-molecule SERS induced in Rhodamin 6G by a silver-island film. Digitized data (dots) from Ref. 3. **c,d** thiophenol self-assembled monolayers on individual dimers of gold nanoparticles. Digitized data (dots) from **c** Fig. 3B and **d** Fig.3A of Ref. 4. The FWHM obtained through fitting with Eq.(8) (red line) are given next to the data.

Dieringer *et al.*[3] noted the small FWHM in their discussion; it was only observed in Raman profiles measured on a single hotspot, whereas ensemble-averaged measurements from multiple hotspots revealed FWHM 100-200 meV. They attributed the narrow resonance reproduced in Fig. S6b to the molecular resonance of the Rhoadmin 6G that got further enhanced by the plasmon coupling. There is no compelling reason to interpret the profile in Fig. S6b as molecular; it might likewise originate from a narrow plasmonic enhancement as suggested by us. Zhu and Crozier[4] assigned their profiles, Fig. 6c and d, to plasmonic resonances. They did not comment on the narrow line width. We note that the resonance at 1.5 eV shown in Fig. 6c and d agrees excellently with the energy expected for an outgoing plasmon-induced resonance in the system studied in Ref. 4. The maxima in the dark-field spectra were close to 900nm (1.38 eV) and the phonon energy 1074cm$^{-1}$ (0.13 eV) yielding a predicted outgoing resonance at 1.51 eV, the same energy as obtained in the fits of Fig. S6c and d. Unfortunately, the incoming resonance was not measured Zhu and Crozier[4], most likely, due to the limited detection range of the Si CCD.

**4. Polarized Raman scattering**

Measuring the polarization dependence of Raman scattering is a delicate task. On first sight one only needs a set of two polarizers to realize any scattering configuration. However, great care needs to be taken that the scattering intensity does not change for reasons related to the measurement setup, not the sample itself. This prohibits, e.g., inserting and removing polarizing elements from the beam path, because they will always absorb and reflect a small fraction of the radiation. Additionally, the throughput of the focusing microscopes and the sensitivity of the spectrometer depend on the polarization of the light. Reich *et al.*[5] discussed how to set up polarization-dependent measurements. Briefly, our Raman setup included a Fresnel rhomb for the incoming laser beam, a λ/2 wave plate in front of the microscope objective, a λ/2 wave plate in the scattered light path, and an analyzer in front of the spectrometer. With this setup any scattering configuration can be realized by turning the polarizing elements, i.e., we do not remove optical elements to switch between the polarization configurations and the light entering the spectrometer is always polarized horizontally.

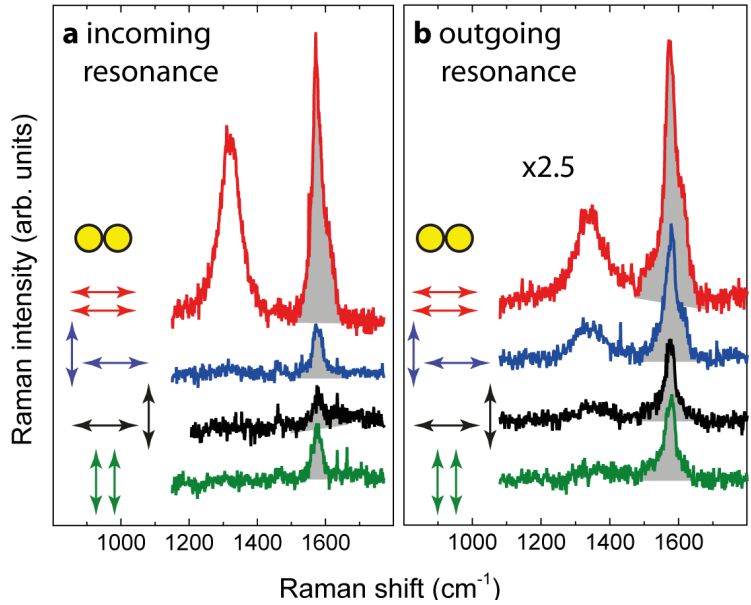

**Supplementary Figure S8:** Polarized Raman spectra as measured at the **a** incoming and **b** outgoing resonance of Dimer1. The yellow circles represent the nanodimer; the arrows indicate the polarization direction of the incoming and outgoing light relative to the dimer axes. All spectra in panel **b** are multiplied by a factor of 2.5.

Supplementary Fig. S8 shows Raman spectra measured on Dimer1 in all four scattering configurations (see labels) at the incoming and outgoing resonances. At both laser energies the intensity is strongest in the $(\|,\|)$ configuration where both incoming and outgoing photon may couple to the nanocavity excitation. The intensity ratio of the $G$ line between $(\|,\|)$ and $(\perp, \perp)$ is five for the incoming resonance in Fig. S6a (the ratio is three for the outgoing resonance in Fig. S6b). This compares well with the ratio between the full plasmonic enhancement versus the enhancement by the delocalized background, which is four at the incoming resonance (two at the outgoing resonance), compare Fig. 2b. We also note that the $D$ and $D'$ modes are absent in the $(\perp, \perp)$ scattering configuration. Both observations mean that the resonance of the nanocavity is a highly polarized hotspot that can only be excited by light polarized parallel to the nanodimer.

At the outgoing resonance, Fig. S6b, the enhancement of the delocalized background is comparable to the enhancement by the nanocavity. To extract the nanocavity enhancement we subtracted the spectrum in (⊥, ⊥) configuration (only background) from the (∥,∥),(⊥,∥), and (∥,⊥) configuration to obtain the spectra in Fig. 5 of the main paper. The plasmonic enhancement by the nanocavity is then analyzed using the quantum mechanical description of plasmon-enhanced Raman scattering.

## 5. Strain analysis through phonon frequency

When the graphene monolayer is transferred onto the plasmonic dimer it adheres to the substrate and the gold nanostructure. The dimer introduces strain in the graphene layer creating a local nanostructure. Strain leads to a shift and splitting of the phonon frequencies. This is an important characteristic of our plasmon-probe system for evaluating plasmon-enhanced Raman scattering, because the plasmonically enhanced modes differ in *frequency* from the background signal, see Refs. 1 and 2 for an in-depth discussion.

We use the measured phonon frequencies in the PERS spectra to obtain the local strain of the graphene on top of the dimer cavity. A general strain in graphene can be divided into a biaxial or hydrostatic component $\varepsilon_h$ that shifts phonon peaks and a uniaxial or shear strain component $\varepsilon_s$ that leads to a splitting of the peaks. The overall frequency shift of a phonon by the hydrostatic strain is given by[1]

$$\Delta\omega = -\omega_0 \gamma_{Gr} \varepsilon_h,$$

where $\omega_0$ is the phonon frequency in the absence of strain and $\gamma_{Gr}$ the mode Grüneisen parameter. The measured *G* line frequencies are 1572cm$^{-1}$ on Dimer1 and 1523cm$^{-1}$ on Dimer2; $\omega_0$=1582cm$^{-1}$ and $\gamma_{Gr}$=1.8. We obtain a strain of 0.4% on Dimer1 and 2.0% on Dimer2. The difference in strain agrees well with the topography observed with AFM, Supplementary Fig. S2. It explains the stronger plasmonic enhancement in Dimer2, where the graphene is closer to the nanocavity hotspot.

## 6. Gloassary of symbols

| Symbol | Meaning |
|---|---|
| **Common Super/Subscripts** | |
| PERS | Plasmon-enhanced Raman scattering |
| LSP(R) | Localized-surface plasmon (resonance) |
| EM | Electromagnetic (referring to the electromagnetic enhancement model) |
| ph ($G$, $2D$, $D$ etc.) | Phonon (specific line: $G$ line, $2D$ line, $D$ line etc.) |
| **Intensities** | |
| $I_G, I_D, I_{2D}$ | Intensity (peak area) of the phonon specified in the subscript |
| $I_{in}, I_{out}$ | Intensity at incoming/outgoing nanocavity resonance (Fig.2) |
| $I_{bg}$ | Intensity of the delocalized plasmonic enhancement (Fig.2) |
| $I_{ph}^{PERS}$ | Calculated intensity using the quantum description of PERS |
| $I_{ph}^{EM}$ | Calculated intensity using the EM model |
| **Matrix elements** | |
| $K_{ph}^{PERS}$ | Raman matrix element calculated for the quantum description of PERS |
| $K_{ph,b/c/d/e}^{PERS}$ | Raman matrix element of a particular scattering pathway (compare Fig.3) |
| $K_{ph}^{EM}$ | Raman matrix element calculated for electromagnetic enhancement |
| $K_{ph}^{Raman}$ | Matrix element for standard Raman scattering |
| $M_{pt-pl}$ ($M_{pl-pt}$) | Matrix element for photon-plasmon (plasmon-photon) coupling |
| $M_{pl-e}$ ($M_{e-pl}$) | Matrix element for plasmon-electron (electron-plasmon) coupling |
| $M_{pt-e}$ ($M_{e-pt}$) | Matrix element for photon-electron (electron-photon) coupling |
| $M_{e-ph}$ | Matrix element for electron-phonon coupling |
| $\tilde{M}$ | Combined matrix element to describe plasmon-probe coupling, see Eq.(3) for a definition |
| **Parameters, independent variables** | |
| $\hbar\omega$ | Photon energy |
| $E_1=\hbar\omega_1$, $E_2=\hbar\omega_2$ | Energy of the incoming (subscript 1) and scattered (2) photon |
| $\hbar\omega_{ph}$ | Phonon energy |
| $E_{LSP}$ | Energy of the LSPR |
| $\gamma_{LSP}$ | FWHM/2 of the LSPR |
| $E_e$ | Electron energy |
| $\gamma_e$ | Electron lifetime |

**References**

[1] S. Heeg, *et al.*, Polarized Plasmonic Enhancement by Au Nanostructures Probed through Raman Scattering of Suspended Graphene, Nano Lett. **13**, 301 (2013).

[2] S. Heeg *et al.*, Strained Graphene as a Local Probe for Plasmon-Enhanced Raman Scattering by Gold Nanostructures, Phys. Stat. Sol. (RRL) **7**, 1067 (2013).

[3] J. A. Dieringer et al., Surface-Enhanced Raman Excitation Spectroscopy of a Single Rhodamine 6G Molecule, JACS 131, 849 (2009).

[4] W. Zhu, K. B. Crozier, Quantum mechanical limit to plasmonic enhancement as observed by surface-enhanced Raman scattering, Nat. Comm. DOI: 10.1038/ncomms6228 (2015).

[5] S. Reich, C. Thomsen, and J. Maultzsch, *Carbon nanotubes: Basic concepts and physical properties* (Wiley-VCH, Weinheim, 2004).